\documentstyle[aps,epsfig]{revtex}                         

\draft \draft
\begin{document}
\twocolumn[\hsize\textwidth\columnwidth\hsize
     \csname @twocolumnfalse\endcsname


\title{Numerical simulation of  interference experiments
       in a local hidden variables model.
       }

\author{W. A. Hofer}
\address{
         Department of Physics and Astronomy,
         University College, Gower Street, London WC1E 6BT, UK}

\maketitle

\begin{abstract}
We present a theoretical model which allows to keep track of all
photons in an interferometer. The model is implemented in a
numerical scheme, and we simulate photon interference measurements
on one, two, four, and eight slits. Measurements are simulated for
the high intensity regime, where we show that our simulations
describe all experimental results so far. With a slightly modified
concept we can also model interference experiments in the low
intensity regime, these experiments have recently been performed
with single molecules. Finally, we predict the result of
polarization measurements, which allow to check the model
experimentally.
\end{abstract}

\pacs{03.65.Vf,03.65.Ta}

\vskip2pc]


\section{Introduction}

The problem of interference has haunted modern physics practically
from the beginning of the twentieth century. Even though
experiments, where a time ordered sequence of single impacts has
been recorded, have only recently been developed
\cite{tonomura89,arndt99}, the thought experiments predicting the
very same outcome ran into considerable conceptual problems much
earlier. For a comprehensive review of the problem see Franco
Selleri's book \cite{selleri92}.

The problem can be stated in the following way. Consider a double
slit interferometer. Only one particle is thought to propagate
from the source to the screen. Therefore the single particle
passes a single slit without any interaction with other particles,
and impinges on the screen. However, even though the particle
cannot be physically influenced by the existence of the second
slit, its impact probability at certain spots of the screen
depends explicitly on the existence of this second slit. How then,
can the particle avoid these spots even though there seems no
physical means to that end? For this reason, we do not have a
single theoretical model at present, which describes the impact of
this particle by a consistent, time ordered sequence of events,
along the particle's path from the source, through the slits and
onto the screen, where all interactions are local. The only model,
which is capable of such an analysis, is non-local. It is based on
David Bohm's theory of hidden variables \cite{philippidis79}.
However, this theoretical model depends explicitly on potentials,
changing the trajectory of the particle between the slits and the
screen, it is therefore unsuitable to describe photon interference
measurements.

We wish to close this gap in the present paper by a model, which
describes the location of every single photon in the
interferometer at every single moment. Moreover, it allows to
recover all the results of the experiments while remaining
perfectly local. The model has been implemented in a numerical
simulation scheme. We perform these simulations mainly for
photons, even though a slightly modified model could also account
for the recent C60 single-slit interference experiments. The aim
of our presentation is to demystify the problem by stating very
clearly and in unambiguous terms how the statistical picture
arises from limited control in the experiments, and how the
physical interactions, which are thought to be deterministic, come
into play. It will be seen that the statistical picture underpins
the usual treatment in classical field theory, e.g. in the scalar
Kirchhoff model, and that it is the very same picture which has
been adopted for quantum mechanical purposes. We make one
modification to the current understanding of this process, i.e.
that the phase of a photon's field does change discontinuously
during the scattering at the slit. And we generally limit the
model to the experimentally verified situations, that is to the
high intensity regime.

However, the low intensity regime, as in electron interference
measurements or recent measurements with molecules, is also
discussed extensively. It is shown that in one case one obtains
identical results, if the single impacts follow a Gaussian
distribution around the actual slits. In the second case, the
molecule interference experiments, our investigation is still
incomplete but we can make it plausible that a similar model to
the one suggested here also accounts for these experiments.

\section{Theoretical model}

In our theoretical description of photon scattering we use
essentially Richard Feynman's path integral method, with a few but
decisive modifications. To incorporate the phase of a photon along
a given path we describe photons as electromagnetic fields of
limited extensions. The compatibility of such a photon model with
quantum mechanics has been shown quite some time ago
\cite{hofer98}. We omit higher Fourier coefficients, which are due
to the confinement of the field, and consider every photon as
essentially monochromatic. The field of the photon possesses an
arbitrary phase $\phi_0$ at the point of emission. This
arbitrariness is due to our lack of knowledge about the individual
photon. To simplify the model we assume that the fields of all
photons are polarized in the same direction, arbitrary
polarization only has the effect that the intensity of the beam
will have to be higher to obtain the same effects. The phase of
the electromagnetic field of the single photon varies along its
optical path from the source to the slit, a difference of $\Delta
\phi = 2 \pi$ describes two separate points with a distance
$\lambda$, the wavelength of the photon. The phase at the slit of
our interferometer is therefore $\phi_1 = \phi_0 + 2\pi
L_1/\lambda$, where $L_1$ is the distance between the source and
the slit.

The scattering of single photons at the slit environment is
described by a statistical model. Since currently no theoretical
treatment of the problem exists, which would allow us to relate
the interaction of the photon with the slit unambiguously to a
direction, into which the photon is scattered, we formalize the
process in the following way:

The phase $\phi_1$ of the photon together with a random phase
$\bar{\phi}_1$ of the slit environment shall trigger a process
which changes the direction of the photon to a scattering angle
$\theta_1$. The angles of all photons measured will comply with a
Gaussian distribution centered around zero. In practice we have
simulated two photons simultaneously, using the phase of the
second photon $\phi_1(2)$ as the input from the slit environment
$\bar{\phi}_1(1)$, and vice versa. Since both phases are random,
this procedure does not couple the photons in any way, it merely
guarantees that the angular distribution after the slit is
Gaussian. The halfwidth of the distribution is one of our input
parameters. The relevant lines in our code are shown in Appendix
A.

Clearly, if the photon changes its direction, then it has
interacted with the slit environment. Since we have no
deterministic model of this process, we also do not know, what the
phase of the photon after the interaction will be. To understand
in more detail the nature of this problem, consider the classical
scalar model of interference experiments. Here, the scalar field
$u({\bf r})$ is thought to be emitted from a point like source.
After scattering at the slits it is adsorbed at the screen.
Throughout the interferometer the field is at every point in phase
with the field at the source. It is thus only by a rigid
connection of field amplitudes via the phase that interference
effects occur at all within this model.

While such a model works perfectly well for laser beams with their
high coherence length and where all photons can be thought to be
in phase, it is less appropriate for ordinary monochromatic light.
These emissions originate from thermal processes, therefore their
phases must generally be random, even though their frequency, e.
g. from the decay of an electron excitation in an atom, can be
sharply defined. Therefore we explore a different possibility in
this paper, and we shall show, that also for a statistical
distribution of initial phases interference effects can be
obtained, provided the scattering process of a single photon at
one slit changes the phase of the photon discontinuously. It goes
without saying that the results are the same, if we use
monochromatic laser light instead of monochromatic light from
emissions of thermally excited atom.

In our model, as indeed in every model which constructs
interferences from uncorrelated photons, the single photon will
have an arbitrary phase at the source. Moreover, the phase of the
photon is not connected to the phase of any other photon.
Therefore, if we adopt the view that the field of the single
photon remains in-phase with the field at its source, we will not
describe any interference effects. We conclude from this fact that
the field cannot remain in phase. Since the photon interacts only
at one point of our system, at the slit of the interferometer, it
can only be this point, where the phase changes.

Here, we make the only truly novel conjecture in this paper: the
phase angle $\phi_1'$, after the scattering process, shall be
proportional to the angle, into which the photon is scattered:

\begin{equation}
\phi_1' = \alpha \theta_1
\end{equation}

For reasons of consistency we choose the constant $\alpha$ equal
to one. The conjecture can then be linked to the description of
field deflection in classical electrodynamics, where a reflection
of a field at a surface leads to a change of its phase by $\pi$
\cite{born70}. Since in classical field theories the phase is
continuous throughout the system, the phase shift is related to
changes throughout the system. In case of arbitrary initial phases
we can only recover this feature, if we set the phase itself equal
to the angle of scattering. In our view this model is a
generalization of classical concepts to the case, where we have
given up the rigid connection of fields throughout the system.

The photons are scattered at a single slit and propagate in a
straight line towards the detector screen. Here, the fields of
individual photons are superimposed and the total field at a point
P(x,y) is given by the expression:

\begin{equation}
E(x,y) = \sum_{i=1}^n cos(\phi_2(i))
\end{equation}

where n is the total number of photons impinging at this point,
and $\phi_2$ is the phase at their moment of impact. The phase is
related to the scattering angle $\theta_1$ by the expression:

\begin{equation}
\phi_2 = \theta_1 + 2 \pi \frac{\sqrt{L^2 + y^2}}{\lambda} \qquad
\frac{y}{L} = tan \theta_1
\end{equation}

\begin{figure}
\begin{center}
\epsfxsize=1.0\hsize \epsfbox{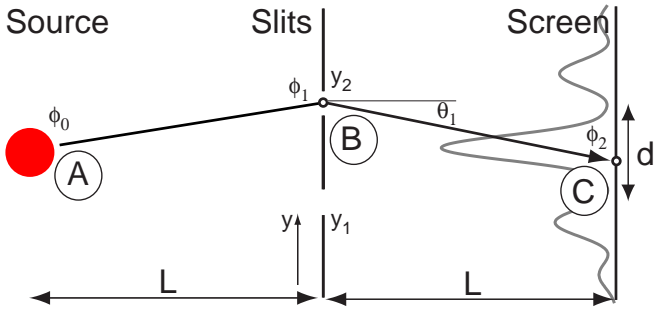}
\end{center}
\caption{Model interferometer in our numerical simulations. (A)
         The photon is emitted from the source with an initial
         phase $\phi_0$. (B) The acquired phase of the photon
         $\phi_1$ and a random phase of the slit environment
         determine the scattering angle $\theta_1$. The phase after
         the scattering process changes discontinuously to
         $\phi_1'$. (C) The photon hits the detector screen
         and interacts with other photons impinging at the same
         location.
         }
\label{fig001}
\end{figure}

Here $L$ is the distance between the plane of the slits and the
plane of the detector screen. The details of the model are
displayed in Fig. \ref{fig001}. The intensity at the point P(x,y)
of the detector screen is consequently:

\begin{equation}
I(x,y) = E(x,y)^2
\end{equation}

The intensity we observe is thus the intensity due to the total
field after superposition of all contributions. It is for this
reason that the model is generally limited to the high intensity
regime. However, we will indicate in the discussion, how a
slightly modified version of the same model can also account for
single molecule interferences. Furthermore, we shall demonstrate
that single electron experiments, where a periodic image is
gradually evolving, can be understood without the occurrence of
any superpositions from separate slits and thus without
interferences proper.

\section{Numerical implementation}

The theoretical model requires a random input at two points of the
interferometer: (i) at the emission of a photon from the source;
and (ii) at the scattering of the photon at the slit environment.
We have used a random number generator for both processes
\cite{random}, mapping the random number [0,1] onto the initial
phase [$0,2\pi$]. The dimensions of the apparatus were chosen
similar to an experimental one. We also chose monochromatic
visible light of $\lambda = 500$nm. The distance $d$ between two
slits of the interferometer equals 0.5 $\mu$m. The distance $L$
between the source and the slit and the slit and the screen in our
simulations was 10mm (see Fig. \ref{fig001}).

The simulation of an experimental run is straightforward. For a
pair of photons, we create the initial phases $\phi_0(1,2)$, and
compute the phase after covering the distance to the slit. Here,
we use the phase of the second photon $\phi_1(2)$ as the random
phase of the slit environment for photon one and vice versa. The
two random inputs are used to create two random outputs, which now
comply with a Gaussian distribution (see Appendix A). This
Gaussian distribution describes the scattering angle of the
photons after the slit. The phase angle of the photons at this
point is set equal to the scattering angle, and both photons
propagate until they hit the screen. We record the points of
impact and the phases at the moment of impact, before simulating
the next pair. In all our simulations we simulated 100000 photons
per slit. We have assumed that the slit is very narrow, in
practice we have always used the same point as the slit position.
The simulation method has been tested for slits of variable width,
but apart from a blurred image due to the variable position of the
photons in the plane of the slits we did not obtain any new
feature compared to the results obtained with a very narrow slit.
Since the slit width is generally unknown we have omitted this
part of the numerical analysis.

To evaluate the photon impacts we have assumed that photons
interact if their distance is less than 10 nm on our detector
screen. This is no limitation of the generality of our model,
since we could always increase the number of photons and thus
allow for higher resolution. The total field at each of these
separate points was calculated and then smoothed with a Gaussian
of 0.1 $\mu$m halfwidth. The plots of the number of photons and
the intensity are shown in units of $I_0$, the maximum intensity,
and $N_0$, the maximum number of impacts.

We have also accounted for a numerical problem in the transition
from single slit diffraction to multiple slit interferometry. In a
single slit, and assuming that the scattering angle is equal to
the phase angle, the intensity distribution of the photons on the
screen is essentially due to the intensity distribution of the
photon fields. All photons, at a given point, will have the same
phase. This is not the case in a multiple slit interferometer.
Here we have to sum up contributions from different slits, and the
total intensity is due to the total electromagnetic field. The
critical points at the screen, in a numerical model, are the
points where the phase difference between separate slits yields
constructive interference, while the phase from one single slit,
or the effect of photon diffraction, yields a minimum of the
field. For a double slit interferometer, the $\cos{\phi_2}$ at the
median point between two slits is zero (because $\phi_2 \approx
\pi/2$), while the interference between the two separate slits is
constructive (because both phases are equal). Since it is largely
a matter of convention in field physics, whether a constant phase
is added in all calculations, we have used the same approach in
this numerical model. For multiple slit simulations we therefore
have increased the final phase $\phi_2$ by a value of $0.2 \pi$
and for every single photon.

\section{Results}

\begin{figure}
\begin{center}
\epsfxsize=1.0\hsize \epsfbox{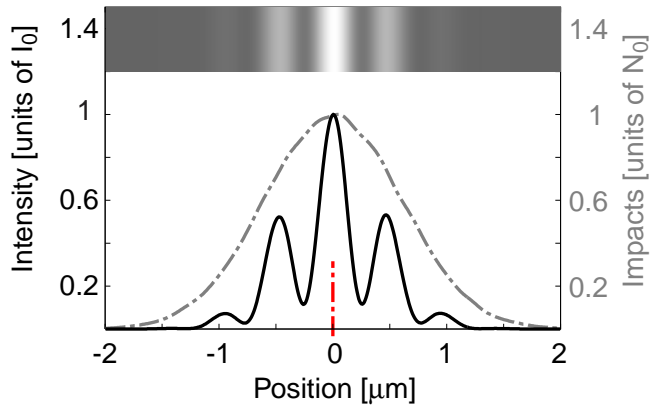}
\end{center}
\caption{Single slit diffraction on very narrow slit. The
         intensity (black curve) shows
         a distinct main peak and two side peaks. The periodic
         feature is due to superposition of the fields of single
         photons, the number density (grey dashed curve) only
         displays the Gaussian distribution of the scattering
         angles. In the filmstrip we show how the image would look
         in a diffractometer.
         }
\label{fig002}
\end{figure}

\begin{figure}
\begin{center}
\epsfxsize=1.0\hsize \epsfbox{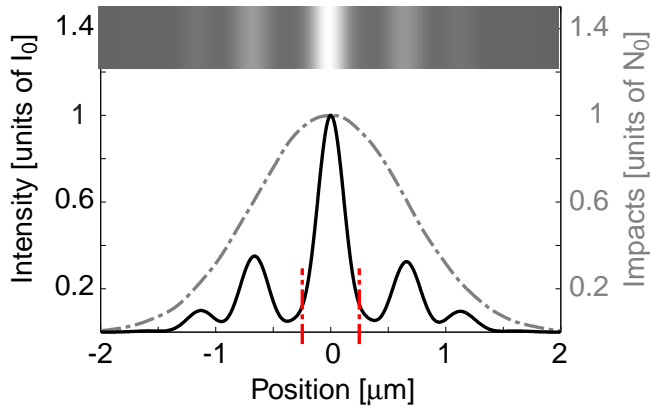}
\end{center}
\caption{Double slit interference. The main peak is between the
         position of the two slits, the plot shows the main
         peak and two side peaks. Note that the maxima at the
         position of the slits have vanished due to destructive
         interference of photons from different slits.
         }
\label{fig003}
\end{figure}

\begin{figure}
\begin{center}
\epsfxsize=1.0\hsize \epsfbox{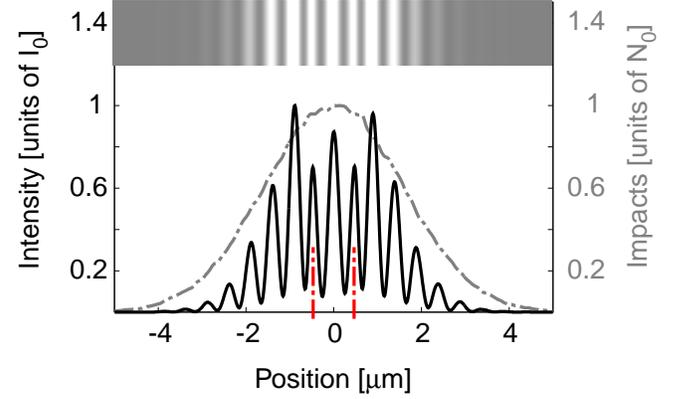}
\end{center}
\caption{Double slit interference with increased slit-distance
        ($d = 2\lambda$) and increased halfwidth ($\sigma = 1.5 \times 10^{-4}$).
         The distance between maxima and minima, which is only due
         to the differences of the optical path for individual photons
         remains constant, the number of observed peaks is substantially
         increased.
         }
\label{fig003a}
\end{figure}

\begin{figure}
\begin{center}
\epsfxsize=1.0\hsize \epsfbox{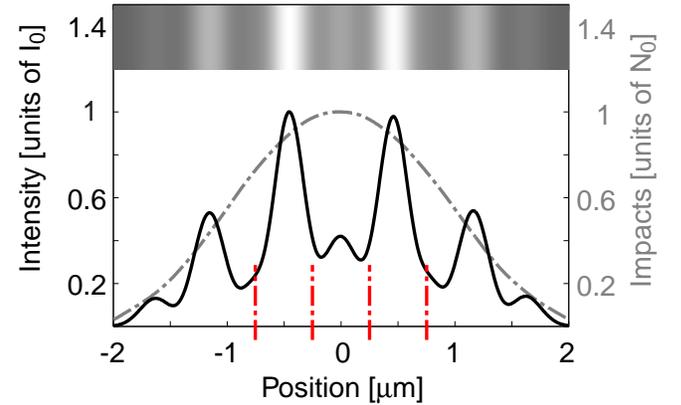}
\end{center}
\caption{Four slit simulation. The image shows two main peaks
         separated by a minor peak of the intensity distribution.
         Comparing with the two slit image the main peak of the
         intensity distribution seems to be reduced. This is due
         to destructive interference of photons from the outer
         slits.
         }
\label{fig004}
\end{figure}

\begin{figure}
\begin{center}
\epsfxsize=1.0\hsize \epsfbox{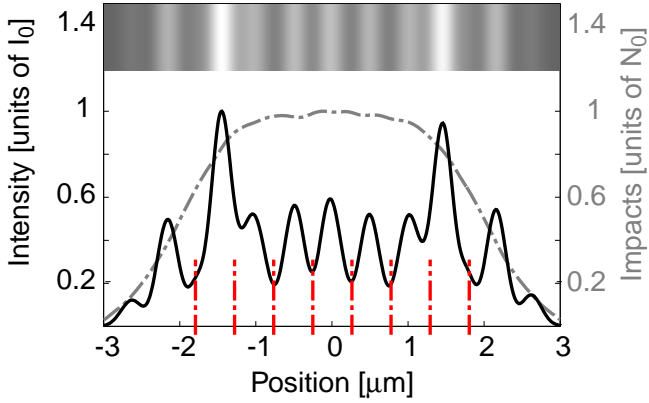}
\end{center}
\caption{Eight slit simulation. The intensity of all five
         inner maxima is roughly equal, the distribution of maxima
         and minima images the slits of the interferometer, even
         though all maxima are between the positions of individual
         slits.
         }
\label{fig005}
\end{figure}

The main results of our simulations are shown in Figs.
\ref{fig002} to \ref{fig005}. In all plots we show the normalized
impact numbers (dash dotted grey curve) as well as the normalized
intensity distribution (black curve). The position of the slits is
indicated by a dash dotted vertical line. We also show a simulated
greyscale image of the detector screen, this filmstrip describes
how the image would look in a real interferometer.

The height of the peaks depends, for a given number of slits,
essentially on the halfwidth $\sigma$ of the scattering angles.
Since this parameter reflects the actual experimental situation,
or rather the unknown scattering process in the slit environment,
we cannot set it to a theoretically required value. For this
reason we can only change it so that the results of our simulation
are in accordance with experimental data. In consequence the
absolute value of a specific peak reflects to a certain extent our
choice of parameters. But the change of the peak height, e.g. if a
double slit and a four slit image are compared, does not depend on
our settings, it is a genuine consequence of the theoretical
assumption that the interference pattern is built up by
superposition of the electromagnetic fields of independent
photons.

Analyzing the difference between a single slit diffraction and
double slit interference (Fig. \ref{fig002} and Fig.
\ref{fig003}), it becomes immediately clear why, in the first
case, the peak is at the position of the slit, while in the second
case it indicates the median position between the two slits. This
is simply due to the interference of photons passing through
different slits, which leads to destructive interference at both
slit positions, thus creating the distinct intensity minima. In
Fig. \ref{fig003a} we show, how the distance between the slits and
the halfwidth effect the intensity distribution. For this image we
have increased the distance between the slits to $2 \lambda$, the
halfwidth $\sigma$ is now 1.5. The number of observed peaks in
this case is substantially increased. The image is equal to the
experimental images obtained in a Young interferometer
\cite{cagnet62}.

The effect of destructive interference leads to an interesting
phenomenon, if the fringes for two and four slits are compared.
Because in the latter case (Fig. \ref{fig004}) the peak at the
center of the screen is reduced, it seems thus that the opening of
two new slits effectively decreases the number of impinging
photons. As the number distribution reveals, this is clearly not
the case. The intensity is reduced due to destructive interference
of the electromagnetic fields carried by each individual photon.
Generally speaking the intensity peaks become more and more
similar with an increasing number of slits, so that in the limit
of infinite slits, or an infinite chain of slits, the intensity
distribution would be purely periodic while the distance between
the peaks is equal to the distance between the slits. This is,
what we actually observe in diffraction experiments with either
photons (x-ray diffraction), or electrons (electron diffraction).

Comparing the distribution of photon impacts with the intensity
distribution, it is clear that the main feature, which creates the
interference pattern, is the loss of intensity at certain points
of the detector screen. These points are due to destructive
interference, in the multiple slit interferometer, or to minima of
the photon fields, in the single slit interferometer. We take this
as a general result of our simulation: periodic intensity
distributions in a process, or interference fringes, are always
due to a loss of intensity at specific points.

Our records of the simulation contain the information about each
individual photon: its trajectory and its phase at the moment of
impact. Information, which according to the current understanding
prohibits the interference of photons from different slits (this
is the principle of the so called which-path or welcher Weg
experiments). Here, we retain all the information and still find
an interference pattern.

\begin{figure}
\begin{center}
\epsfxsize=1.0\hsize \epsfbox{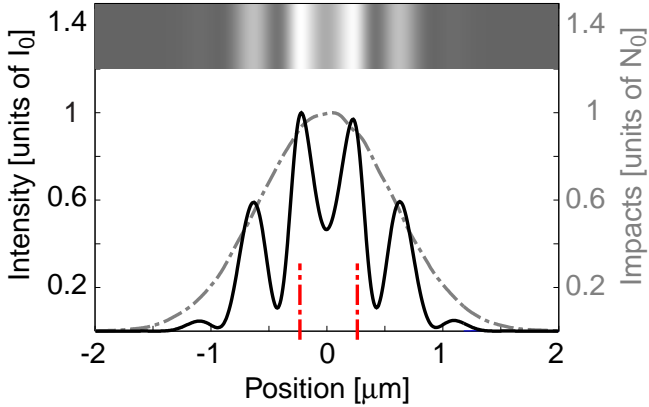}
\end{center}
\caption{Welcher Weg experiment with two slit interferometer. The
         photons are either horizontally (slit 1) or vertically
         (slit 2) polarized. Due to the vector characteristics of
         the electromagnetic fields only photons from the same slit
         interfere for with each other. The double slit image,
         obtained if all photons possess equal polarization, is
         lost (compare the image Fig. \ref{fig003}).
         }
\label{fig006}
\end{figure}

This is due to the fact that the vanishing interference pattern in
a Welcher Weg experiment is no consequence of the information
about the path of the photon, but, in the way the experiment is
often implemented, due to the vanishing superposition of the
photon's electromagnetic fields. In practice, such an experiment
can be carried out with polarized photons in a two slit
interferometer, if photons of horizontal (vertical) polarization
pass only through the first (second) slit. If we simulate this
situation, then we have to account for the vector characteristics
of photon polarization. This, in turn, means that the
electromagnetic field of the horizontally polarized photons is
perpendicular to the electromagnetic field of the vertically
polarized photons. And then the superposition of the fields occurs
independently, and in two dimension, the two directions
perpendicular to the photon's vector of motion. The result of this
simulation is shown in Fig. \ref{fig006}. The two single slits are
still clearly visible, the maxima, furthermore, are at the
position of the slits, clearly photons from one slit have not
interacted with photons from the other slit. The typical two slit
interference pattern can be made to reappear by inserting a
diagonal polarizer into the path of all photons: since all photons
then have the same vector of polarization, the superposition
occurs again only in one dimension and the interference pattern is
recovered.

\section{Discussion}

The local model has two key ingredients: a local scattering event
at a single slit, where the photon is scattered into a new
direction, and a local superposition at the focal plane of the
interferometer, where the electromagnetic fields of single photons
are superimposed. The second feature, the local superposition of
fields, is also the main feature of classical models of
interference. It is, as David Deutsch remarked in his book
\cite{deutsch94}, also the main issue in quantum mechanical
models, even though not every photon can be thought to be real in
every situation. One has to conclude, therefore, that no model
exists, which predicts interference effects from the scattering
events alone. This would imply a non-local connection between the
individual slits. Theoretically speaking we have no reason to
believe that such a model is physically meaningful. Therefore,
even given the present results, the conceptual problem  in the low
intensity regime still seems to prevail. To our knowledge these
low intensity experiments so far have only been performed with
electrons, and C60 molecules \cite{tonomura89,arndt99}. We are
currently investigating the experiments with C60 molecules and the
emerging picture is roughly the following: due to its high energy
the molecule possesses a considerable dipole moment, which
oscillates along its path. This oscillating moment plays the same
role as the periodic electromagnetic field of a photon. In the
slit environment the molecule is scattered into a new direction
due to interaction between the molecule and the atoms of the slit.
The scattering angle depends on the phase of the molecular dipole
moment. The molecule can only be detected after ionization. Since
the ionization process will not be successful in every case, not
all molecules will actually be detected. And if the success rate
of ionization depends on the dipole moment of the molecule, then
the angle of scattering is linked to the probability of detection.
We will present the exact numerical results for this experiments
in a future publication.

\begin{figure}
\begin{center}
\epsfxsize=1.0\hsize \epsfbox{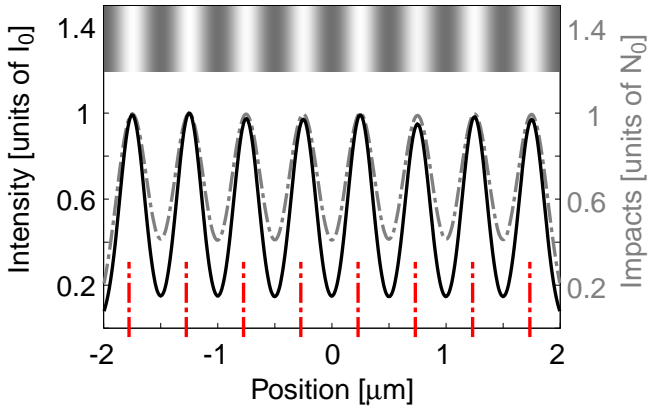}
\end{center}
\caption{Eight slit simulation with reduced angular distribution
         after scattering. The impact regions of photons from
         different slits overlap only in a minor way. In
         consequence, the periodic distribution of intensity
         is no indication of interference effects, even though
         the image looks similar to an eight slit interferometer
         image (compare Fig. \ref{fig005}).
         }
\label{fig007}
\end{figure}

The impact of single electrons gradually building up the
interference pattern of the statistical theory, as in Tonomura's
1989 experiments, seems to be the strongest proof, that a single
electron undergoes interference on the screen, even though there
is no other electron present, which could do the trick. In the
view of the Many Worlds interpretation of quantum mechanics
\cite{dewitt73}, the electron in this case interacts with an
electron from a parallel universe, a 'shadow' electron. As already
mentioned, there is so far no unambiguous proof that the same
behavior can be seen for single photons: photon interference
measurements have only be performed in the high intensity regime.
This makes electron interferences all the more important. And, if
these experiments are correct, then it could be seen as a
demonstration of the validity of the Many Worlds interpretation.
However, an identical result can be obtained, if the single
electrons do not interact with 'shadow' electrons, but if the
angular distribution of scattered electrons after the slit system
is very narrow. So narrow, in fact, that the overlap between the
impact of electrons from separate slits is close to zero. We have
simulated such a measurement with photons, an identical result
could be obtained for electrons if the lengthscale is reduced by
three orders of magnitude. As seen in Fig. \ref{fig007}, the peaks
of the number distribution coincide with the peaks of the
intensity distribution. This is fundamentally different from the
previous cases, where the intensity peaks were always between the
positions of single slit. It is, in short, the proof that there is
no interference at work under these conditions. The periodic
distribution at the screen is merely an image of the periodic
arrangement of slits of the interferometer.

\begin{figure}
\begin{center}
\epsfxsize=1.0\hsize \epsfbox{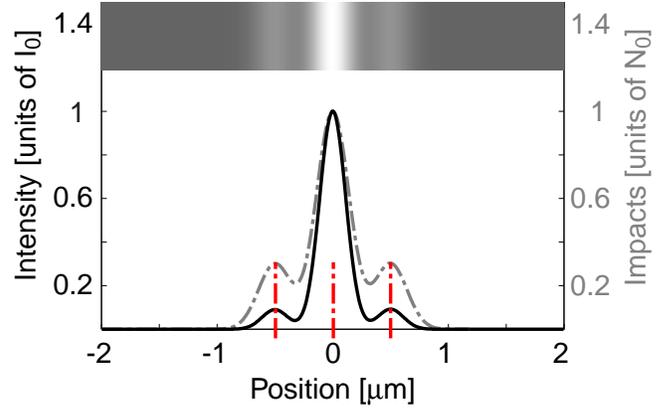}
\end{center}
\caption{Simulation of photon impacts in a three slit system with
         reduced beam width and reduced angular distribution
         after scattering. The impact count reproduces all the
         features of the experiments with single C60 molecules.
         Also in this case the distribution of intensity
         is no indication of interference effects.
         }
\label{fig008}
\end{figure}

Following this line of research, it is also possible that the
experiments with single C60 molecules \cite{arndt99} are no
indication of any phase effects at all, but simply due to a
specific choice of experimental parameters. If we assume that the
initial molecular beam before the slit system is wider than the
distance between the single slits, so that molecules will pass
through the central slit and two side-slits, then the image on the
screen will show one main peak between two smaller peaks. Apart
from determining the absolute position of the image in relation to
the central slit, there is no way to decide, whether we deal with
interference, or just a statistical distribution of molecule
impacts due to their passing through the slit system. We have
simulated this situation by counting the impacts of single
photons, if 40\% of the photons pass through a side-slit. The
Gaussian in this case was assumed to be narrow enough to retain
the side peaks as distinct features. This can essentially be
obtained in an experiment with molecular beams by making the
molecular velocity high enough. The result of this simulation is
shown in Fig. \ref{fig008}. It can be seen that the essential
features of the experimental image - one main peak, two side
peaks, the visibility of the minima very low - are recovered if
the parameters are suitably chosen. Also in this situation we do
not deal with interference, but merely with the impact of
particles scattered at the slit system.

This simple model can easily be checked experimentally. Since the
occurrence of ''interference fringes'' depends on the statistical
spread and thus the velocity of the C60 molecules, the fringes
should disappear, if the velocity decreases below a certain
threshold. In fact, the predictions of the model can be tested by
varying the velocity. We predict that the {\em visibility} depends
on molecular velocity, while in interference experiments proper
the {\em position} of the maxima and minima depends on the
velocity. If the intensity distribution is only due to kinetic
effects, then the positions will not be altered by a variation of
velocity.

Let us go back to our analysis of the double-slit interferometer
in the introduction and to the problem, how photons can avoid
regions of low intensity on the detector screen. It becomes clear
now, that this is actually not the case. The number density of
photons on the screen in all cases of physical interference is
more or less Gaussian. The periodicity of the intensity, the
observed interference pattern, is only due to interference between
the fields of individual photons. The number density, therefore,
is generally not related to the intensity distribution. It appears
thus that quantum mechanics, which arrives at an oscillating
number density (the probability distribution $|\psi|^2$) to
account for the intensity in interference experiments, gives a
misleading and even wrong account of the actual physical
situation.

\section{Experimental tests}

The conjecture used to construct the local model also for a random
distribution of photon phases has interesting consequences, which
can in principle be checked experimentally. Consider an experiment
with polarized monochromatic light. Then after the slit the
direction into which a photon is scattered is related to the phase
after the interaction with the slit environment. Therefore photons
at different points of the detector screen will have different
phases. The difference between e.g. the first and second maximum
in a double slit interferometer (see Fig. \ref{fig003a})
corresponds to a phase difference of

\begin{equation}
\Delta \phi \approx \pi
\end{equation}

For polarized photons, this difference can be measured with an
optical modulator, since the phase at the front end of the
modulator determines the rotation of the field vector
\cite{weihs98,hofer01b}. Subjecting photons after the modulator to
a polarization measurement should yield a different result for
photons impinging at different points of the detector screen.

\section*{Acknowledgements}

A number of people have shared my interest in interference
phenomena during the last year, most notably R. Stadler, who
contributed more to this paper then he is willing to take credit
for. I am also obliged to J. Gavartin, he provided the code for
the numerical scattering model. And finally I have to thank the
UCL HiPerSPACE center for providing the computer infrastructure,
which made the simulations possible.

\section*{Appendix A}

The actual code fragment, which transforms a uniform distribution
[0,1] into a Gaussian distribution $\exp(-x^2/2)$ includes only 15
lines. The main features are: first create a random number PHASE1,
add the phase from the propagation to the slit, and throw away
multiples of one (or of $2 \pi$): this random number is now
transformed from the interval [0,1] (PHI1) to the interval [-1,1]
(V1):

\vspace{0.5cm}
   10   PHASE1 = RANDOM()

        PHI1 = PHASE1 + Z

        K = PHI1

        PHI1 = PHI1 - FLOAT(K)

        V1 = 2. * PHI1 - 1.
\vspace{0.5cm}

Now do the same for a second number PHASE2, which results in V2:

\vspace{0.5cm}
        PHASE2 = RANDOM()

        PHI2 = PHASE2 + Z

        K = PHI2

        PHI2 = PHI2 - FLOAT(K)

        V2 = 2. * PHI2 - 1.
\vspace{0.5cm}

Calculate the radius and see, whether it is inside the halfwidth
of the desired Gaussian: if it is, then compute the final values,
if it is not, then choose two new random numbers:

\vspace{0.5cm}
        RADIUS = V1*V1+V2*V2

        IF (RADIUS.GT.1.0) GOTO 10

        FACTOR = SQRT(-2.*LOG(RADIUS)/RADIUS)

        GAUSS1 = V1*FACTOR

        GAUSS2 = V2*FACTOR
\vspace{0.5cm}

These two variables, GAUSS1 and GAUSS2, which comply with a
Gaussian of halfwidth one, are then multiplied by our input
parameter $\sigma$, the chosen Gaussian halfwidth of our angular
distribution. In the simulations the halfwidth was set to $6
\times 10^{-5}$.

\newpage

\end{document}